\documentclass[letterpaper, 10pt, journal]{ieeeconf} %
\IEEEoverridecommandlockouts                              
\usepackage{mathtools}
\usepackage{amsfonts,amssymb}
\usepackage{arydshln,amsmath}
\usepackage{graphicx}
\usepackage{calc,cases}
\usepackage{float}
\usepackage{color,soul}
\usepackage[utf8]{inputenc}
\usepackage{etoolbox}
\AtBeginEnvironment{bmatrix}{\setlength{\arraycolsep}{2pt}}

\definecolor{magenta}{cmyk}{0.5, 1, 0, 0}
\definecolor{greenedu}{cmyk}{1, 0, 1, 0.1}
\definecolor{cyanedu}{cmyk}{1, 0, 0, 0.1}

\newcommand\xqed[1]{%
  \leavevmode\unskip\penalty9999 \hbox{}\nobreak\hfill
  \quad\hbox{#1}}
\newcommand\demo{\xqed{$\triangle$}}

\newcommand{\G}{G_0(q)}

\newcommand{\Tde}{T_d(q,\eta)}
\newcommand{\nmp}{NMP transmission zero}

\newcommand{\beq}{\begin{equation}}
\newcommand{\eeq}{\end{equation}}
\newcommand{\beqna}{\begin{eqnarray}}
\newcommand{\eeqna}{\end{eqnarray}}

\newtheorem{theorem}{Theorem}
\newtheorem{assumption}{Assumption}
\newtheorem{remark}{Remark}

\title{\LARGE \bf
Optimal Controller Identification for multivariable non-minimum phase systems
}

\author{D.D. Huff, L. Campestrini, G.R. Gonçalves da Silva and A.S. Bazanella
\thanks{This work was supported in part by the Coordena\c{c}\~{a}o de Aperfei\c{c}oamento de Pessoal de N\'{i}vel Superior - Brazil (CAPES) - Finance Code 001, and by the Conselho Nacional de Desenvolvimento Cient\'{i}fico e Tecnol\'{o}gico - Brazil (CNPq).}
\thanks{D.D. Huff is with GIPSA-Lab, Universit\'{e} Grenoble Alpes, Grenoble, France. Email: {\tt\footnotesize daniel.denardi-huff@gipsa-lab.fr}}%
\thanks{L. Campestrini and A.S. Bazanella are with Programa de P\'{o}s-Gradua\c{c}\~{a}o em Engenharia El\'{e}trica, Universidade Federal do Rio Grande do Sul, Porto Alegre, Brazil. Emails: {\tt\footnotesize \{luciola,bazanella\}@ufrgs.br}}%
\thanks{G.R. Gonçalves da Silva is with Department of Electrical Engineering, IPCOS, Boxtel, The Netherlands.
        Email: {\tt\footnotesize gustavo.rodrigues.gs@gmail.com}}
}
\begin{document}

\maketitle
\thispagestyle{empty}
\pagestyle{empty}

\begin{abstract}
This paper extends the formulation of a data-driven control method -- the Optimal Controller Identification (OCI) -- to cope with non-minimum phase (NMP) systems, without \textit{a priori} knowledge of the NMP transmission zero, i.e. without obtaining a prior model of the plant -- as in any data-driven approach. The method is based on the Model Reference paradigm, in which the desired closed-loop performance is specified by means of a closed-loop transfer function -- the reference model. Considering a convenient parametrization of the latter and a flexible performance criterion, it is possible to identify the NMP transmission zeros of the plant along with the optimal controller parameters, as it will be shown. Both diagonal and block-triangular reference model structures are treated in detail. Simulation examples show the effectiveness of the proposed approach.
\end{abstract}


\section{Introduction} 

Unlike model-based control, data-driven control methods are not based on the knowledge of a process model. Instead, they aim at designing the parameters of a fixed-structure controller through the use of a reasonably large batch of input-output data collected from the process \cite{Bazanella:Campestrini:Eckhard:2012}. 
The classical data-driven control design methods are typically based on the Model Reference (MR) paradigm, in which the desired closed-loop performance is specified by means of a closed-loop transfer function -- the \textit{reference model}. The Optimal Controller Identification (OCI) method \cite{Campestrini:Eckhard:Bazanella:Gevers:2017,Huff:Campestrini:Goncalves:Bazanella:2019,Huff:Goncalves:Campestrini:2018}, which is the focus of this work, fits into this class.

The non-minimum phase (NMP) systems issue in Model Reference approaches is a well-known problem in control literature and it is no different in data-driven methods. The problem arises when the system presents transmission zeros (cf. the definition in \cite{MacFarlane&Karcanias:1976}) outside the unit circle (mainly those that come from the discretization of right-half-plane zeros in continuous-time domain). These zeros tend to be canceled out by the controller when they are not present in the reference model, since one cannot move the zero location with feedback, leading to an internally unstable closed loop \cite{Skogestad:2007}. For multivariable systems the situation is even more complex, because the transmission zeros also have input and output directions associated to them. These directions have implications on output performance and allowable reference model choices \cite{Goncalves:Campestrini:Bazanella:2016}.

Identification and inclusion of these zeros in the reference model is now an established and safer solution whether in direct adaptive control \cite{astrom1980,Ioannou} or data-driven control designs \cite{Campestrini:Eckhard:Gevers:Bazanella:2011, Goncalves:Campestrini:Bazanella:2016, Goncalves:Campestrini:Bazanella:2018}. In \cite{Goncalves:Campestrini:Bazanella:2018}, a solution is proposed to the NMP issue in the multivariable case using the VRFT (Virtual Reference Feedback Tuning) data-driven method with a flexible criterion, where identification of the NMP transmission zeros is achieved along with the controller parameters in an iterative least-squares procedure. The solution, however, presents a drawback: the reference model has to be diagonal (with identical elements along the diagonal in a first step and possible retuning with different elements). Thus, the NMP transmission zero effect appears in all outputs. A solution for moving the effect of the NMP transmission zero to one output with VRFT is presented in \cite{Goncalves:Campestrini:Bazanella:2016}. However, the output direction of the zero must be known \textit{a priori} in order to define the reference model. In other words, a model for the plant must be previously identified to apply the method.

In this work, we adapt the multivariable formulation of the OCI method \cite{Huff:Campestrini:Goncalves:Bazanella:2019} in order to alleviate the aforementioned constraints and show that a design where the NMP effect is moved to a single output can also be achieved without identifying the plant. Both diagonal and block-triangular reference model structures are considered. Similar to the solution proposed in \cite{Campestrini:Eckhard:Gevers:Bazanella:2011,Goncalves:Campestrini:Bazanella:2018} for the design with VRFT, we propose a flexible criterion for the OCI such that the NMP transmission zeros and the optimal controller
can be simultaneously identified.

The paper is organized as follows. Section~\ref{sec:problem} presents definitions and problem description. The flexible formulation of the OCI method is presented in Section~\ref{sec:OCI}. Section~\ref{sec:nmp} reviews the issue regarding NMP systems and discusses in more detail how to apply the OCI method for such systems. Section~\ref{sec:caveat} explains how to deal with the case of diagonal reference models, which requires special care. Section~\ref{sec:examples} presents some simulation examples. The paper ends with some concluding remarks.

\section{Preliminaries}\label{sec:problem}
Consider a linear time-invariant discrete-time multivariable process
\begin{equation}\label{eq:y(t)}
y(t) = G_0(q)u(t) + H_0(q)w(t),
\end{equation}
where $q$ is the forward-shift operator, $u(t)$ and $y(t)$ are $n$-vectors representing the
process' input and output, respectively, and $w(t)$ is a sequence of independent random $n$-dimensional vectors with zero mean values and covariance matrix  $E [w(t)w^T(t)] = \Lambda$. 
The transfer matrix $G_0(q)$ and the noise model $H_0(q)$  are square $n\times n$ 
matrices whose elements are proper rational transfer functions and $H_0(\infty)=I$. 

The design task is to tune the parameter vector $P\in\mathcal D_P\subset\mathbb{R}^{n_P}$ of a linear time-invariant controller $ C(q,P)$ in order to achieve a desired closed-loop response, expressed in terms of the closed-loop poles. We assume that this controller belongs to a given user-specified  controller class $\mathcal{C}$ such that all elements of the loop transfer matrix
$L(q)=G_0(q)C(q,P)$ have positive relative degree for all $C(q,P) \in \mathcal{C}$.
The control action $u(t)$ can be written as
\begin{equation}
	u(t) = C(q,P)e(t) = C(q,P)(r(t) - y(t)),
	\label{eq:u(t)}
\end{equation}
where $r(t)$ is the reference signal, which is assumed to be quasi-stationary and uncorrelated with the noise $w(t)$,
that is, $\bar E \left[  r(t)w^T(t-\tau) \right]   = 0,~ \forall \tau $, where \cite{Ljung:1999}:
$$ \bar{E}[f(t)] \triangleq \lim_{N\rightarrow\infty} \frac{1}{N}\sum_{t=1}^{N} E[f(t)]$$
with $E[\cdot ]$ denoting expectation. 
The system \eqref{eq:y(t)}-\eqref{eq:u(t)} in closed loop becomes
\begin{equation}
\label{eq:ycl} y(t,P) = T(q,P)r(t) + (I-T(q,P))v(t),
\end{equation}
\begin{equation}
\label{eq:defm} T(q,P) = [I + G_0(q)C(q,P)]^{-1}G_0(q)C(q,P),
\end{equation}
where $v(t)=H_0(q)w(t)$ and the dependence on the controller parameter vector $P$ is now made explicit in the output signal $y(t, P)$. 

The controller class $\mathcal{C}$ is defined as
$$ \mathcal{C}= \left\{ C(q,P) : P \in \mathcal D_P \subset \mathbb{R}^{n_P} \right\},$$
where the structure of the controller to be designed is
\begin{equation}
C(q,P) =
\scalebox{0.95}{$\left[\begin{array}{cccc} C_{11}(q,\rho_{11}) & C_{12}(q,\rho_{12}) & \cdots & C_{1n}(q,\rho_{1n}) \\
\vdots & \vdots & \ddots & \vdots \\ C_{n1}(q,\rho_{n1}) & C_{n2}(q,\rho_{n2}) & \cdots & C_{nn}(q,\rho_{nn})  \end{array}\right]$}
\label{eq:controller}
\end{equation}
and $P=[\rho_{11}^T ~~\rho_{12}^T~~\ldots ~~\rho_{n1}^T~~\ldots ~~\rho_{nn}^T]^T$. We assume that $C(q,P)$ is invertible for all $P\in\mathcal D_P$.

In the case studies presented in this paper, we will consider PID controllers of the form
\begin{equation}
C_{ij}(q,\rho_{ij}) = 
\dfrac{a_{ij}q^2+b_{ij}q+c_{ij}}{q(q-1)}
\label{eq:pid_struct}
\end{equation}
where $\rho_{ij}=[a_{ij}~~ b_{ij} ~~c_{ij}]^T$ and the derivative pole is fixed at zero.

In the Model Reference approach to the design, the closed-loop performance is specified through a rational transfer matrix $T_d(q,\eta)$, known as the reference model. While the poles of $T_d(q,\eta)$ are typically fixed according to the desired closed-loop behavior, the numerators of the elements of $T_d(q,\eta)$ can optionally be parametrized by a vector $\eta\in\mathcal D_\eta\subset\mathbb{R}^{n_\eta}$ and let free to vary in order to increase the flexibility of the method, as it will become clearer in Section~\ref{sec:nmp} (see also \cite{Campestrini:Eckhard:Gevers:Bazanella:2011} and \cite{Goncalves:Campestrini:Bazanella:2018}). The controller parameters are then tuned as the solution of the problem:
\begin{equation}\label{minJMR}
	P^{MR} = \underset{P\in\mathcal D_P}{\text{arg min}} ~ \underset{\eta\in\mathcal D_\eta}{\text{min}} ~ J^{MR}(P,\eta),
\end{equation}
\begin{equation}\label{eq:JMR}
	J^{MR}(P,\eta) \triangleq \frac{1}{N} \sum_{t=1}^{N} ||(T_d(q,\eta) - T(q,P))r(t)||^2_2,
\end{equation}
where $r(t)$ is the reference signal of interest and $N$ is the time horizon.

The \textit{ideal controller} $C_d(q,\eta)$ is the one that allows the closed-loop system behavior
to match exactly the one prescribed by $T_d(q,\eta)$:
\begin{equation}
\label{eq:contideal} C_d(q,\eta) = G_0^{-1}(q)T_d(q,\eta)[I-T_d(q,\eta)]^{-1}.
\end{equation}

For our further analysis, we will sometimes consider the following assumption:
\begin{assumption} \underline{Matching condition}\label{asp:fullorder}
\vspace{0.1cm}
\begin{center}
$\exists~ \theta_d =  [P_d^T\ \eta_d^T]^T \in \mathcal D_P\times\mathcal D_\eta\mbox{ such that}$
\end{center}
\vspace{0.1cm}
\begin{equation}\label{eq:matching_eta}
C(q,P_d) = C_d(q,\eta_d).
\end{equation}
\end{assumption}

Notice that when this assumption holds $T_d(q,\eta_d)=T(q,P_d)$ and $J^{MR}(P_d,\eta_d)=0$. That is, the closed-loop response with the controller $C(q,P_d)$ matches the one prescribed by $T_d(q,\eta_d)$. The specific value $\eta_d$ assumed by the parameter $\eta$ is not as crucial as the value assumed by $P$, since $\eta$ does not influence the poles of the reference model $T_d(q,\eta)$.

\section{Optimal Controller Identification}\label{sec:OCI}

Using the concept of the ideal controller, it is possible to turn the control design problem into 
an identification problem of the system, without using a model for the process. For that purpose we will adapt the OCI method \cite{Campestrini:Eckhard:Bazanella:Gevers:2017,Huff:Goncalves:Campestrini:2018,Huff:Campestrini:Goncalves:Bazanella:2019} to the case where the reference model is parametrized, following the idea presented in \cite{Goncalves:Campestrini:Bazanella:2018} for the VRFT method. This formulation is specially suitable for NMP processes.
 
The core idea is to rewrite the input-output  system \eqref{eq:y(t)} in terms
of the ideal controller $C_d(q,\eta)$, which is done by inverting the relation (\ref{eq:contideal}), i.e.
\beq 
G_0(q) = T_d(q,\eta)\left(I - T_d(q,\eta)\right)^{-1}C_d^{-1}(q,\eta).\label{eq:G0}
\eeq
Then a model for the plant can be re-written as a function of the reference model and of the chosen controller structure:
\begin{align}
G(q,\theta) &\triangleq   T_d(q,\eta)\left(I - T_d(q,\eta)\right)^{-1}C^{-1}(q,P)\nonumber\\
                                                &\triangleq L_d(q,\eta)C^{-1}(q,P) \label{eq:G}
\end{align}
where $\theta = [P^T\ \eta^T]^T \in \mathcal D_\theta = \mathcal D_P\times \mathcal D_\eta \subset \mathbb{R}^{n_P+n_\eta}$ and $\mathcal D_\theta$ is a compact and connected set. We assume that $G(q,\theta)$ is causal for all $\theta\in\mathcal D_\theta$. The task will be to fit a model $G(q,\hat\theta)$ to $G_0(q)$ which, with the formulation \eqref{eq:G}, amounts to obtaining an estimate $C(q,\hat P)$ of the ideal controller $C_d(q,\hat\eta)$ (where $\hat\eta$ is the value identified for $\eta$).

More specifically, from $N$ measured input-output data, the parameter vector estimate $\hat{\theta}_N = \left[\hat{P}_N^T\ \ \hat{\eta}_N^T\right]^T$ is defined as \cite{Huff:Campestrini:Goncalves:Bazanella:2019}:
\begin{equation}
	\hat{\theta}_N = \underset{\theta\in\mathcal D_\theta}{\text{arg min}}\ V_N(\theta)\\
	 \label{eq:V}
\end{equation}
where
\begin{equation}
	V_N(\theta) \triangleq \frac{1}{N}\sum_{t=1}^N\left\|\epsilon(t,\theta)\right\|_2^2, \label{eq:V1b}
\end{equation}
$\epsilon(t,\theta)$ is the prediction error
\begin{equation}
	\epsilon(t,\theta) \overset{\Delta}{=} y(t) - \hat{y}(t\mid t-1,\theta)
	\label{eq:epsilon}
\end{equation}
and 
\begin{equation}
	\hat{y}(t\mid t-1,\theta) =G(q,\theta)u(t) =L_d(q,\eta)C^{-1}(q,P)u(t)
	\label{eq:preditor}
\end{equation}
is the one-step ahead predictor related to the output error (OE) model structure:
\begin{equation}
y(t,\theta) = G(q,\theta)u(t) + w(t).
\label{eq:y(t)OE}
\end{equation}

Instead of minimizing $J^{MR}(P,\eta)$, which depends on the unknown plant $G_0(q)$, the design is made by minimizing the 
cost function $V_N(\theta)$, which is purely data-dependent and no model of the plant $G_0(q)$ is used.
Since the estimation of the controller has been transformed into a PE identification problem, all properties of
PE identification theory apply. Specifically, the estimate in \eqref{eq:V} converges with probability 1 for $N\rightarrow\infty$ to the
vector $\theta^* = [P^{*T}\ \eta^{*T}]^T$ defined as follows \cite{Ljung:1999}:
\begin{equation}
	\hat{\theta}_N \rightarrow \theta^* = \underset{\theta\in\mathcal D_\theta}{\text{arg min}}\ \bar{V}(\theta)
\end{equation}
where
\begin{equation}
	\bar{V}(\theta) = \bar{E}\left\|\epsilon(t,\theta)\right\|_2^2.
\label{eq:V2}
\end{equation}

Moreover, if Assumption~\ref{asp:fullorder} is satisfied and an informative enough data set is collected in open loop, then, for $N \rightarrow \infty$ \cite{Ljung:1999}:
\begin{align}
	G(q,\hat{\theta}_N) \rightarrow G_0(q),
\end{align}
that is, 
\begin{align}\label{eq:Cp_Cd}
	C(q,\hat P_N) - C_d(q,\hat\eta_N) \rightarrow 0.
\end{align}
In this case the ideal controller will be \textit{consistently} identified.

A similar consistency result exists if data are collected in closed loop and a (sufficiently generic) noise model structure $H(q,\theta) $ is used to identify $H_0(q)$. The reader is refered to \cite{Huff:Campestrini:Goncalves:Bazanella:2019} for the details.

\section{Non-minimum Phase Systems}\label{sec:nmp}

In this section, we review the issue regarding NMP transmission zeros and see in more detail how to apply the previously presented method to NMP systems. Two different structures for the reference model $T_d(q,\eta)$ are exploited.

If the ideal controller $C_d(q,\eta)$ were put in the control loop, the objective function $J^{MR}(P,\eta)$ would evaluate to zero, providing the desired input-output performance. However, it follows from \eqref{eq:contideal} that the plant's (transmission) zeros turn into poles of the ideal controller, which will result in internal instability for plants that possess NMP zeros. That is, unless the NMP zeros appearing in the denominator of \eqref{eq:contideal} are canceled by a proper choice of the reference model, as specified in the following theorem.

\begin{theorem}\cite{Havre&Skogestad:1996} \label{teo:interznmp}
If $\G$ has an \nmp{} at $z_{nm}$ with output direction $y_{z_{nm}}$, then for internal stability of the feedback system with the ideal controller, the following constraint must apply:
\begin{equation}
y_{z_{nm}}^HT_d(z_{nm},\eta)=0.
\label{eq:inter_znmp}
\end{equation}
\end{theorem} 

Theorem \ref{teo:interznmp} states that in order to obtain internal stability,
the reference model $T_d(q,\eta)$ must have at least the same \nmp{s} of $G_0(q)$ in the same output directions. 
So if the system has non-minimum phase transmission zeros the user needs to know at least
their location (just like in the SISO case \cite{Bazanella:Campestrini:Eckhard:2012}) in order to choose a reference model satisfying \eqref{eq:inter_znmp}. 

Any \textit{a priori} knowledge about the \nmp{} location and output direction (obtained via system identification, for example) may be helpful when choosing the reference model. However, data-driven methods are advantageous because they do not require a system identification step. The idea then is to obtain both the \nmp{} and the controller parameters in the same algorithm. That is the reason why we suggest to parametrize the reference model by means of a vector $\eta$, which will be identified along with $P$. 

\subsection{Parametrization of the reference model}\label{sec:parametrization}

We now give two examples of reference models that satisfy the constraints pointed out above and show how they can be parametrized using the vector $\eta$.

\subsubsection{Diagonal reference model}
Consider first a diagonal reference model:
\begin{equation}\label{eq:estrut_diag}
\mbox{\small $T_d(q)=\begin{bmatrix}
T_{d_{11}}(q) &\ldots     &0      \\
\vdots &\ddots     &\vdots     \\
0 &\ldots     &T_{d_{nn}}(q)      
\end{bmatrix}$}.
\end{equation}
A diagonal reference model can be considered to be the easier and most common choice, since the user defines only the performance requirements for each loop and also aims at complete decoupling. Following the guidelines of \cite{Goncalves:Bazanella:Campestrini:2019}, we propose for each element (assuming, for instance, that the plant possesses only one \nmp{} $z_{nm}$):
\begin{equation}
T_{d_{jj}}(q) = \dfrac{\frac{(1-p_{1,jj})(1-p_{2,jj})}{1-z_{nm}}(q-z_{nm})}{(q-p_{1,jj})(q-p_{2,jj})},
\label{eq:tdjj}
\end{equation}
for $j=1,\ldots,n$, where constraint~\eqref{eq:inter_znmp} is satisfied since $T_d(z_{nm})=0$. The poles of~\eqref{eq:tdjj} must be chosen according to performance criteria. Instead of the second-order transfer function~\eqref{eq:tdjj}, it would also be possible to consider higher order models. 
Since the \nmp{} $z_{nm}$ is, in principle, unknown in practice, we can parametrize~\eqref{eq:tdjj} with $\eta$ in order to automatically identify $z_{nm}$ when applying the OCI method:
\begin{equation}\label{eq:tdjjeta0}
T_{d_{jj}}(q,\eta) = \frac{\eta_{1,jj}q+\eta_{2,jj}}{(q-p_{1,jj})(q-p_{2,jj})}.
\end{equation}

In the case studies presented in Section~\ref{sec:examples}, we consider the static gain constraint $T_d(1,\eta)=I,~\forall\eta\in\mathcal D_\eta$ (for constant references). Note that in this case~\eqref{eq:tdjjeta0} is equivalent to
\begin{equation}\label{eq:tdjjeta}
T_{d_{jj}}(q,\eta) = 
\scalebox{1.2}{$\frac{\eta_{1,jj}q+(1-p_{1,jj})(1-p_{2,jj})-\eta_{1,jj}}{(q-p_{1,jj})(q-p_{2,jj})}$}
\end{equation} 
since $\eta_{1,jj}$ and $\eta_{2,jj}$ are related by $T_{d_{jj}}(1,\eta)$ = 1. This direct substitution avoids the introduction of equality constraints when solving the optimization problem~\eqref{eq:V}.

\begin{remark}
As explained in \cite{Goncalves:Bazanella:Campestrini:2019}, in some cases, it may be convenient to specify a relative degree $\nu_j>1$ for $T_{d_{jj}}(q,\eta)$. This can be accomplished, for instance, as follows:
\begin{equation*}
T_{d_{jj}}(q,\eta) = \frac{\eta_{1,jj}q+\eta_{2,jj}}{q^{\nu_j-1}(q-p_{1,jj})(q-p_{2,jj})}.
\end{equation*}
\end{remark}

\subsubsection{Block-triangular reference model}

Consider now the case where the reference model is not diagonal but satisfies~\eqref{eq:inter_znmp}. In this case, the NMP transmission zero $z_{nm}$ \textit{of the reference model} must have an output direction $y_{z_{nm}}$ equal to that of the process, but its input direction can be different from the process zero input direction. One special case is when the reference model has the block-triangular structure below, which is also considered in \cite{Goncalves:Campestrini:Bazanella:2016}:
\begin{equation}
\mbox{\small $T_d(q)=\begin{bmatrix}
T_{d_{11}}(q) &0      &\ldots      &0      &\ldots     &0      \\
0      &T_{d_{22}}(q) &\ldots      &0      &\ldots     &0      \\
\vdots &\vdots &\ddots &\vdots &     &\vdots \\
T_{d_{k1}}(q) &T_{d_{k2}}(q) &\ldots &T_{d_{kk}}(q) &\ldots     &T_{d_{kn}}(q) \\
\vdots      &\vdots      &      &\vdots      &\ddots     &\vdots      \\
0      &0      &\ldots      &0      &\ldots     &T_{d_{nn}}(q) 
\end{bmatrix}$.}
\label{eq:estrut_movezero}
\end{equation}
This structure allows a design where one can move the effect of the NMP transmission zero to the $k$-th output \cite{Goncalves:Campestrini:Bazanella:2016}. Such a choice of reference model is of interest in cases where the $k$-th output is considered to be less important than the others, so the inverse response (due to the NMP zero) and other possible inconvenient effects should be moved to this particular output.

The elements $T_{d_{jj}}(q), j\neq k$, can be chosen according to the desired performance using models of first or second-order, for instance. To illustrate the key ideas, let us consider first-order models:
\begin{equation}
T_{d_{jj}}(q) = \frac{1-p_j}{q-p_j}.
\label{eq:tdjjeta2}
\end{equation}

On the other hand, the element $T_{d_{kk}}(q)$ must have the NMP transmission zero in order to satisfy~\eqref{eq:inter_znmp}. Let us consider the parametrization~\eqref{eq:tdjjeta} for it (with $j$ replaced by $k$), where the poles are fixed according to performance criteria.

The other elements of row $k$ may be chosen as suggested in \cite{Goncalves:Campestrini:Bazanella:2016}:
\begin{align}\label{eq:tdkj} 
T_{d_{kj}}(q) &= \frac{K_j(q-1)(q-z_{kj})}{(q-p_j)(q-p_{1,kk})(q-p_{2,kk})} \\
&\triangleq K_j(q-z_{kj})\overline{T}_{d_{kj}}(q),\nonumber
\end{align}
where $T_{d_{kj}}(q)$ has the poles of both $T_{d_{jj}}(q)$ and $T_{d_{kk}}(q)$ and the variables $K_j$ and $z_{kj}$ are (dependent) degrees of freedom that can be used to satisfy $y_{z_{nm}}^HT_d(z_{nm})=0$. If the location and output direction of the NMP transmission zero are known, this can be achieved as follows: choose a value for $K_j$ and compute $z_{kj}$ through
\begin{equation}
z_{kj} = z_{nm} + \frac{y_j T_{d_{jj}}(z_{nm})}{y_k K_j\overline{T}_{d_{kj}}(z_{nm})}
\label{eq:z_kj}
\end{equation}
where $y_j$ denotes the $j$-th component of $y_{z_{nm}}^H$ and it is assumed that $y_k\neq 0$.

On the other hand, this procedure cannot be applied if the location and output direction of the NMP transmission zero are not known \textit{a priori}. In this case, we suggest to parametrize~\eqref{eq:tdkj} as:
\begin{equation}\label{eq:tdkjeta}
T_{d_{kj}}(q,\eta) = \frac{(\eta_{1,kj}q+\eta_{2,kj})(q-1)}{(q-p_j)(q-p_{1,kk})(q-p_{2,kk})}.
\end{equation}

\section{Diagonal reference models: a caveat}\label{sec:caveat}

Since the approach of Section~\ref{sec:OCI} is based on the OE model structure~\eqref{eq:y(t)OE}, it cannot be employed without modifications in some situations. Besides the case where the plant~\eqref{eq:y(t)} is unstable, special care must be taken when applying the proposed method to NMP systems with a diagonal reference model, as explained below.

If the reference model has more \nmp{s} than the process (or with a higher multiplicity), then those extra zeros must come from the controller\footnote{Actually, it is also possible to choose a non-diagonal reference model with more \nmp{s} than the plant, but when $T_d(q)$ is diagonal the chances are higher because each element will possess the zero (see \eqref{eq:tdjj}).} (cf. \cite{Skogestad:2007}). The problem is that \eqref{eq:G} depends on the inverse of $C(q,P)$, which will be unstable if $C(q,P)$ has \nmp{s}, causing the OE identification procedure of Section~\ref{sec:OCI} to fail. Let us see an example to illustrate this:
\newline \newline
\textbf{Example 1. }Consider a process described by
\begin{equation}
G_0(q)=
\begin{bmatrix}
\frac{(q-0.7)}{(q-0.9)(q-0.8)} & \frac{2}{(q-0.8)} \\[4pt]
\frac{1.25}{(q-0.8)} & \frac{1.5}{(q-0.8)}\\[4pt]
\end{bmatrix},
\label{eq:sis_znmp}
\end{equation}
which has an \nmp{} at $q=1.2$ with output direction $y_{z_{nm}}=[-0.6~~0.8]^T$. We choose the following reference model, which satisfies Assumption~\ref{asp:fullorder} for the PID controller class:
\begin{equation}
	T_d(q)=\begin{bmatrix}
		\frac{-0.4(q-1.2)}{(q-0.6)(q-0.8)}&0\\
		0&\frac{-0.4(q-1.2)}{(q-0.6)(q-0.8)}
	\end{bmatrix}.
	\label{eq:Tddiag0}
\end{equation}
Notice that \eqref{eq:Tddiag0} has the same \nmp{} as \eqref{eq:sis_znmp} but with a higher multiplicity. The ideal controller in this case is given by the expression
\begin{equation}
\hspace*{-0.4cm}C_d(q) =
	\begin{bmatrix}
		\frac{0.6(q-0.9)(q-0.8)}{q(q-1)}&\frac{-0.8(q-0.9)(q-0.8)}{q(q-1)}\\
		\frac{-0.5(q-0.9)(q-0.8)}{q(q-1)}&\frac{0.4(q-0.8)(q-0.7)}{q(q-1)}
	\end{bmatrix}
\end{equation}
whose inverse is unstable:
\begin{equation}
\hspace*{-0.4cm}
\small C_d^{-1}(q) = \begin{bmatrix}
		\frac{-2.5(q-0.7)(q-1)q}{(q-0.8)(q-0.9)(q-1.2)}&\frac{-5.0(q-1)q}{(q-0.8)(q-1.2)}\\
		\frac{-3.1(q-1)q}{(q-0.8)(q-1.2)}&\frac{-3.8(q-1)q}{(q-0.8)(q-1.2)}
	\end{bmatrix}.
\end{equation}
That is, if $C(q,P)$ is parametrized according to~\eqref{eq:controller}-\eqref{eq:pid_struct}, $G(q,\theta)$ will be unstable in a neighborhood of the solution $\theta=\theta_d$ (even if $G(q,\theta_d)=G_0(q)$ is stable). Thus, the optimization problem~\eqref{eq:V} cannot be solved in practice.

\demo

To overcome the difficulties presented above, we propose an approach similar to the one of \cite{Forssell:Ljung:2000}, where an OE model structure is adapted to identify an unstable system. 
Let us rewrite \eqref{eq:G} as
\begin{equation}
G(q,\theta) = L_d(q,\eta)\frac{N(q,P)}{D(q,P)}
\label{eq:G2}
\end{equation}
where $D(q,P)$ is the least common denominator of all entries of $C^{-1}(q,P)$ and $N(q,P)$ is the matrix of the corresponding numerators. Notice that $N(q,P)$ and $D(q,P)$ are uniquely defined up to a scalar factor, whose choice is arbitrary. For simplicity, we assume that the leading coefficient of $D(q,P)$ does not vanish for all $P\in\mathcal D_P$ and that the filter $L_d(q,\eta)$ is stable for all $\eta\in\mathcal D_\eta$ (except for the pole it typically contains at one, because it will be canceled out in the product $L_d(q,\eta)C^{-1}(q,P)$ if $C(q,P)$ has integral action).

If $C(q,P)$ is a PI controller for instance
\begin{equation}\label{eq:PI}
C(q,P)=
\frac{1}{q-1}\begin{bmatrix}
a_{11}q+b_{11} & a_{12}q+b_{12}\\
a_{21}q+b_{21} & a_{22}q+b_{22}
\end{bmatrix}
\end{equation}
then we have:
\begin{align}
&D(q,P) = (a_{11}a_{22}-a_{12}a_{21})q^2+ \nonumber \\&  (a_{11}b_{22}+a_{22}b_{11}-a_{12}b_{21}-a_{21}b_{12})q + b_{11}b_{22}-b_{12}b_{21} \label{eq:DPpol}\\
&N(q,P) = (q-1)\begin{bmatrix}
a_{22}q+b_{22} & -(a_{12}q+b_{12}) \\
-(a_{21}q+b_{21}) & a_{11}q+b_{11}
\end{bmatrix}.\label{eq:N}
\end{align}

Let $n_D$ be the degree of $D(q,P)$. Let us also define $\delta(P) \triangleq [d_0(P)\ d_1(P)\ \ldots \ d_{n_D}(P)]^T$ as the vector of coefficients of $D(q,P)$. In the example~\eqref{eq:PI}-\eqref{eq:N}, $n_D=2$ and
\begin{equation}\label{eq:deltaexemplo}
\small
\hspace*{-0.3cm}\delta(P) =
\begin{bmatrix}
d_0(P)\\d_1(P)\\d_2(P)
\end{bmatrix}=
\begin{bmatrix}  
a_{11}a_{22}-a_{12}a_{21} \\ 
a_{11}b_{22}+a_{22}b_{11}-a_{12}b_{21}-a_{21}b_{12} \\
b_{11}b_{22}-b_{12}b_{21}
\end{bmatrix}.
\end{equation}

With a slight abuse of notation, one then gets 
\begin{align}
D(q,\delta) = d_0q^{n_{D}} + d_1q^{n_{D}-1}+\ldots+d_{n_{D}} = D(q,P). \label{eq:D}
\end{align}

Let us also define the polynomials $D_S(q,\delta_S)$ and $D_U(q,\delta_U)$ as the stable and unstable factors of $D(q,\delta)$, respectively, with $D_U(q,\delta_U)$ monic. That is:
\begin{align}
\label{eq:DsDa}
D(q,\delta) &= D_S(q,\delta_S)D_U(q,\delta_U)\\
\label{eq:Ds}D_S(q,\delta_S) &\triangleq s_0q^{n_S} + s_1q^{n_S-1}+\ldots+s_{n_S}\\
\label{eq:Da}D_U(q,\delta_U) &\triangleq q^{n_U} + u_1q^{n_U-1}+\ldots+u_{n_U}
\end{align}
where $\delta_S\triangleq[s_0\ s_1\ \ldots\ s_{n_S}]$ and $\delta_U\triangleq[u_1\ \ldots\ u_{n_U}]$. Note that $n_S+n_U=n_D$, but the degree of each polynomial depends on the number of unstable roots of $D(q,\delta)$. Moreover, note that $\delta=\delta(P)$ is an \textit{explicit} function of $P$ (see the expressions in~\eqref{eq:deltaexemplo}, for instance), while $\delta_S=\delta_S(P)$ and $\delta_U=\delta_U(P)$ are \textit{implicit} functions of $P$, i.e. they cannot be computed in a straightforward manner using the elements of $P$. We assume that $D(q,P)$ does not have any root on the unit circle for all $P\in\mathcal D_P$, otherwise the decomposition \eqref{eq:DsDa}-\eqref{eq:Da} is not well defined. If $\mathcal D_P$ is a connected set, this also means that the values $n_U$ and $n_S$ are constant over $\mathcal D_P$. 

Furthermore, let
\begin{equation}
D_U^*(q,\delta_U) \triangleq u_{n_U}q^{n_U} + u_{n_U-1}q^{n_U-1}+\ldots+1
\label{eq:Da2}
\end{equation}
be the polynomial whose roots are equal to the inverse of the roots of $D_U(q,\delta_U)$.

To solve the instability problem illustrated by Example~1, we pre-multiply the prediction error $\epsilon(t,\theta)$ in \eqref{eq:V1b} by the all-pass filter
\begin{equation}
F(q,P) \triangleq \frac{D_U(q,\delta_U(P))}{D_U^*(q,\delta_U(P))}I.
\label{eq:F}
\end{equation}
Using~\eqref{eq:epsilon} and \eqref{eq:preditor}, one has
\begin{align}
\small
&\epsilon_F(t,\theta) \triangleq  F(q,P)\epsilon(t,\theta)\\
&= \frac{D_U(q,\delta_U)}{D_U^*(q,\delta_U)}y(t)-L_d(q,\eta)\frac{N(q,P)}{D_S(q,\delta_S)D_U^*(q,\delta_U)}u(t)
\label{eq:Fepsilon}
\end{align}
where all the filters are stable and the dependence of $\delta_S$ and $\delta_U$ on $P$ is omitted. Thus, instead of minimizing~\eqref{eq:V1b} we minimize:
\begin{equation}
V_{N,F}(\theta) = \frac{1}{N}\sum_{t=1}^N\left\|\epsilon_F(t,\theta)\right\|_2^2.
\label{eq:VNF}
\end{equation}

This procedure provides a consistent estimate of the ideal controller if Assumption~\ref{asp:fullorder} is satisfied and an informative enough data set is collected in open loop, just like the standard OE structure \cite{Forssell:Ljung:2000}.

\subsection{Gradient calculations}\label{sec:gradient}

In order to minimize \eqref{eq:VNF}, we apply at a first step the steepest-descent method and at a second step the Levenberg-Marquardt regularization procedure, as it is done in \cite{Huff:Campestrini:Goncalves:Bazanella:2019}. Both methods require the first-order partial derivatives of the prediction error $\epsilon_F(t,\theta)$ with respect to $\theta$, which can be calculated in the usual manner \cite{Ljung:1999}. However, the procedure discussed above requires special attention, since $\epsilon_F(t,\theta)$ depends on $\theta$ not only through $P$ and $\eta$ but also indirectly through
\beq
\bar\delta(P) \triangleq
\begin{bmatrix} \delta_S(P)\\ \delta_U(P) \end{bmatrix}
\eeq
which is an implicit function of $P$. The goal of this section is to show how to obtain the jacobian of $\epsilon_F(t,\theta)$ with respect to $\theta$, that is, the following $n\times (n_P+n_\eta)$ matrix:
\begin{align}
\small
&\dfrac{\partial \epsilon_F(t,\theta)}{\partial \theta} \triangleq 
\begin{bmatrix}
\dfrac{\partial \epsilon_F(t,\theta)}{\partial P} & \dfrac{\partial \epsilon_F(t,\theta)}{\partial \eta}
\end{bmatrix} \nonumber\\
&\triangleq 
\begin{bmatrix}
\dfrac{\partial \epsilon_F(t,\theta)}{\partial P_1} & \cdots & \dfrac{\partial \epsilon_F(t,\theta)}{\partial P_{n_P}} & \dfrac{\partial \epsilon_F(t,\theta)}{\partial \eta_1} & \cdots & \dfrac{\partial \epsilon_F(t,\theta)}{\partial \eta_{n_\eta}}
\end{bmatrix}
\label{eq:epsilon_derivada}
\end{align}
where $\dfrac{\partial \epsilon_F(t,\theta)}{\partial P_k}$ denotes the partial derivative of $\epsilon_F(t,\theta)$ with respect to the $k$-th element of $P$ and similarly for $\eta$.

Let us define
\beq
 \Psi \triangleq \begin{bmatrix} \theta \\ \bar\delta \end{bmatrix}=\begin{bmatrix} P \\ \eta \\ \bar\delta \end{bmatrix}.
\eeq

Consider now the jacobian of $\epsilon_F(t,\Psi)\equiv\epsilon_F(t,\theta)$ with respect to $\Psi$: 
\beq
\dfrac{\partial \epsilon_F(t,\Psi)}{\partial \Psi} \triangleq 
\begin{bmatrix}
\dfrac{\partial \epsilon_F(t,\Psi)}{\partial P} & \dfrac{\partial \epsilon_F(t,\Psi)}{\partial \eta}
& \dfrac{\partial \epsilon_F(t,\Psi)}{\partial \bar\delta}
\end{bmatrix}.
\eeq

Matrix~\eqref{eq:epsilon_derivada} can then be obtained through the chain rule:
\beq
\dfrac{\partial \epsilon_F(t,\theta)}{\partial \theta} = \dfrac{\partial \epsilon_F(t,\Psi)}{\partial \Psi}
\begin{bmatrix}
I_{n_P} & 0\\
0 & I_{n_\eta}\\
\dfrac{\partial \bar\delta}{\partial P} & 0
\end{bmatrix},
\eeq
where $I_k$ denotes the $k\times k$ identity matrix and
\beq
\frac{\partial \bar\delta}{\partial P} =
\begin{bmatrix}
\dfrac{\partial \delta_S}{\partial P} \\[0.2cm] \dfrac{\partial \delta_U}{\partial P}
\end{bmatrix}\in\mathbb{R}^{n_S+1+n_U\times n_P}
\eeq
is the jacobian of $\bar\delta(P)$ with respect to $P$. We still have to show how to obtain $\dfrac{\partial \bar\delta}{\partial P}$. From~\eqref{eq:D} and~\eqref{eq:DsDa}-\eqref{eq:Da}, we know that $P$ and $\bar\delta$ satisfy the following equality:
\begin{equation}
\delta(P) = h(\bar\delta),\\
\label{eq:restricao}\end{equation}
where the function $h(\cdot)$ is known explicitly and depends on the convolution of $\delta_S$ and $\delta_U$. Consider, for instance, the example \eqref{eq:PI}-\eqref{eq:deltaexemplo} (where $n_D=2$) with $n_S=n_U=1$. In this case, one has
\begin{align*}
\delta(P) &= \begin{bmatrix}  
a_{11}a_{22}-a_{12}a_{21} \\ 
a_{11}b_{22}+a_{22}b_{11}-a_{12}b_{21}-a_{21}b_{12} \\
b_{11}b_{22}-b_{12}b_{21}
\end{bmatrix}\\
&= \begin{bmatrix}
s_0 \\ s_0u_1+s_1 \\ s_1u_1
\end{bmatrix} = h(\bar\delta).
\end{align*}

Applying the implicit function theorem, we then get
\begin{equation}
\dfrac{\partial \bar\delta}{\partial P} = \left[ \frac{\partial h(\bar\delta)}{\partial \bar\delta}\right]^{-1}\dfrac{\partial \delta(P)}{\partial P}
\end{equation}
where
$$
\dfrac{\partial h(\bar\delta)}{\partial \bar\delta} =
\begin{bmatrix}1  & 0   & 0 \\
							u_1 & 1   & s_0\\
							0   & u_1 & s_1
\end{bmatrix}
$$
is the jacobian of $h(\bar\delta)$ with respect to $\bar\delta$ and similarly for $\dfrac{\partial \delta(P)}{\partial P}$. Both $\dfrac{\partial h(\bar\delta)}{\partial \bar\delta}$ and $\dfrac{\partial \delta(P)}{\partial P}$ are obtained by direct calculation.

\begin{remark}
The matrix $\dfrac{\partial h(\bar\delta)}{\partial \bar\delta}$ has a special structure. Consider, for instance, that $n_S=3$ and $n_U=2$. Then, one obtains:
\begin{equation*}
\dfrac{\partial h(\bar\delta)}{\partial \bar\delta}=
\setlength{\dashlinegap}{2pt}
\left[\begin{array}{c:ccccc}
1 		& 0 		& 0 		& 0 	& 0 	   	&0		\\ \hdashline[4pt/2pt]
u_1 & 1 		& 0 		& 0 	& s_0 	&0		\\
u_2 & u_1 & 1 		& 0 	& s_1 	&s_0\\
0 		& u_2 & u_1 & 1 	& s_2		&s_1\\
0 		& 0 		& u_2 & u_1 & s_3	&s_2\\
0 		& 0 		& 0 		& u_2 & 0			&s_3
\end{array}
\right]
\end{equation*}
where the lower right corner corresponds to the Sylvester matrix associated to the polynomials $D_U(q,\delta_U)$ and $D_S(q,\delta_S)$, which is known to be nonsingular if and only if these polynomials do not share any root (which is the case). Thus, it follows that $\dfrac{\partial h(\bar\delta)}{\partial \bar\delta}$ is also nonsingular, as required by the implicit function theorem. 
\end{remark}

\section{Simulated Examples}\label{sec:examples}

This section presents the application of the proposed methodolgy to a simulated plant.
All simulation results consider the plant given by~\eqref{eq:sis_znmp}. Data are collected from the system in closed loop with the proportional controller $C_0(q) = 0.5I$. The reference is set as a pseudo-random binary sequence (PRBS) of amplitude~1, fundamental period of 20 samples and duration of 1260 samples.

\subsection{Diagonal reference model}

Consider the following reference model
\begin{equation}
\hspace*{-0.3cm}T_d(q,\eta)=
	\begin{bmatrix}
		\frac{\eta_{1,11}q+0.08-\eta_{1,11}}{(q-0.6)(q-0.8)}&0\\
		0&\frac{\eta_{1,22}q+0.12-\eta_{1,22}}{(q-0.6)(q-0.7)}
	\end{bmatrix}
	\label{eq:Td1eta}
\end{equation}
where we are specifying a response at output~2 faster than at output~1.

The designed PID controller is given by
\begin{equation}
\hspace*{-0.4cm}\scalebox{0.8}{$C(q,\hat P_N)=$}
	\begin{bmatrix}
		\frac{0.60(q-0.897)(q-0.813)}{q(q-1)}&\frac{-1.08(q-0.896)(q-0.831)}{q(q-1)}\\
		\frac{-0.50(q-0.903)(q-0.808)}{q(q-1)}&\frac{0.56(q-0.810)(q-0.730)}{q(q-1)}
	\end{bmatrix}
	\label{eq:C1}
\end{equation}
which corresponds to the following identified reference model:
\begin{equation}\label{eq:Tdi1}
\hspace*{-0.4cm}T_d(q,\hat \eta_N)=
	\begin{bmatrix}
		\frac{-0.393(q-1.204)}{(q-0.6)(q-0.8)}&0\\
		0&\frac{-0.492(q-1.244)}{(q-0.6)(q-0.7)}
	\end{bmatrix}.
\end{equation}
It is important to notice that, since Assumption~\ref{asp:fullorder} is not satisfied, both transfer functions above depend on the controller $C_0(q)$ and on the spectrum of the reference signal employed to collect the batch of data used to design the controller.

\begin{figure}[htb]%
\centering%
\includegraphics[width=0.5\textwidth]{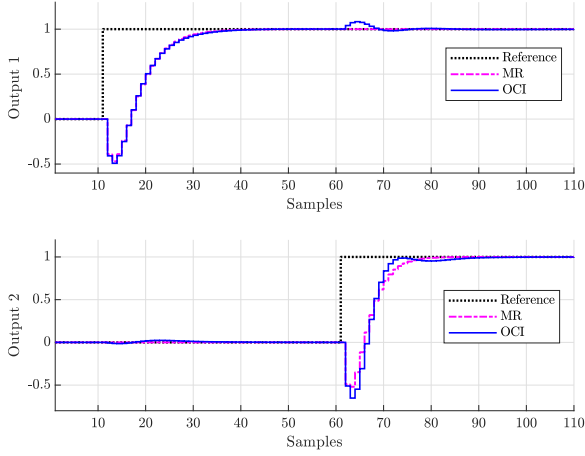}%
\caption{Closed-loop response with controller \eqref{eq:C1}.}
\label{fig:DIAGNideal}%
\end{figure}

The resulting closed-loop response to step references is shown in Fig.~\ref{fig:DIAGNideal}, where we have also plotted the reference model obtained (denoted by MR). The performance measure~\eqref{eq:JMR} is $J^{MR}(\hat P_N,\hat \eta_N)=2\times 10^{-3}$. The obtained response is very similar to the one defined by the identified diagonal $T_d(q,\hat \eta_N)$, where the second output is practically decoupled from the first output and a small coupling is obtained in the first output. On the other hand, the inverse response is present in the step response of both outputs, as expected.


\subsection{Block-triangular reference model}

\begin{figure}[htb]%
\includegraphics[width=0.5\textwidth]{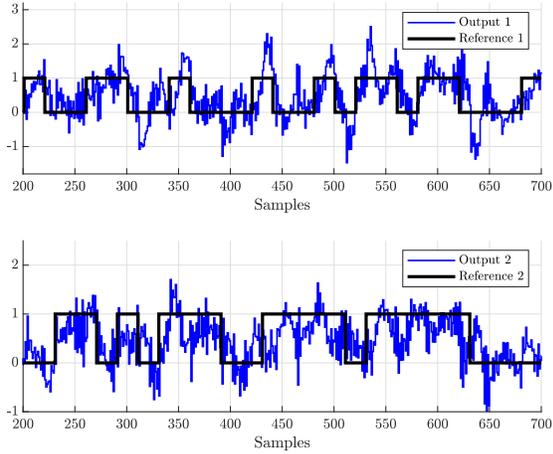}%
\caption{Part of the batch of closed-loop data collected from the plant with $H_0(q)=I$.}
\label{fig:batch_outputs}%
\end{figure}

Consider the case where the NMP effect is to be moved to output $1$. We then select the following parametrized reference model:
\begin{equation} \label{eq:td2eta}
\Tde = \begin{bmatrix}
 \frac{\eta_{1,11}q+0.08-\eta_{1,11}}{(q-0.8)(q-0.6)} & \frac{(\eta_{1,12}q+\eta_{2,12})(q-1)}{(q-0.8)(q-0.6)(q-0.75)} \\[4pt]
 0 & \frac{0.25}{(q-0.75)} \\[4pt]
\end{bmatrix}.
\end{equation}
The transfer function $\Tde$ satisfies Assumption~\ref{asp:fullorder} for the PID controller class with 
\begin{equation}
T_d(q,\eta_d) = \begin{bmatrix}
 \frac{-0.4(q-1.2)}{(q-0.8)(q-0.6)} & \frac{q-1}{(q-0.6)(q-0.75)} \\[4pt]
 0 & \frac{0.25}{(q-0.75)} \\[4pt]
\end{bmatrix}.
\label{eq:Td2}
\end{equation}
The collected data in closed loop with $C_0(q)$ are now corrupted by noise, corresponding to ${H_0(q)=I}$ and 
$$\Lambda = \begin{bmatrix}0.04 & 0\\ 0 & 0.02\end{bmatrix}$$ in~\eqref{eq:y(t)}. The signal-to-noise ratio (SNR) is of approximately $9$~dB at both outputs. Fig.~\ref{fig:batch_outputs} shows part of the batch of data.

The designed PID controller is given by
\begin{equation}
\hspace*{-0.4cm}\scalebox{0.8}{$C(q,\hat P_N)=$}
	\begin{bmatrix}
		\frac{0.62(q-0.889)(q-0.822)}{q(q-1)}&\frac{-1.09(q-0.909)(q-0.794)}{q(q-1)}\\
		\frac{-0.52(q-0.877)(q-0.838)}{q(q-1)}&\frac{1.08(q-0.797)(q-0.770)}{q(q-1)}
	\end{bmatrix}
	\label{eq:C2}
\end{equation}
which corresponds to the following identified reference model:
\begin{equation}
T_d(q,\hat \eta_N)=
	\begin{bmatrix}
		\frac{-0.379(q-1.211)}{(q-0.6)(q-0.8)}&\frac{(q-1)(q-0.813)}{(q-0.6)(q-0.75)(q-0.8)}\\
		0&\frac{0.25}{q-0.75}
	\end{bmatrix}.
	\label{eq:Tdi2}
\end{equation}

\begin{figure}[htb]%
\includegraphics[width=0.5\textwidth]{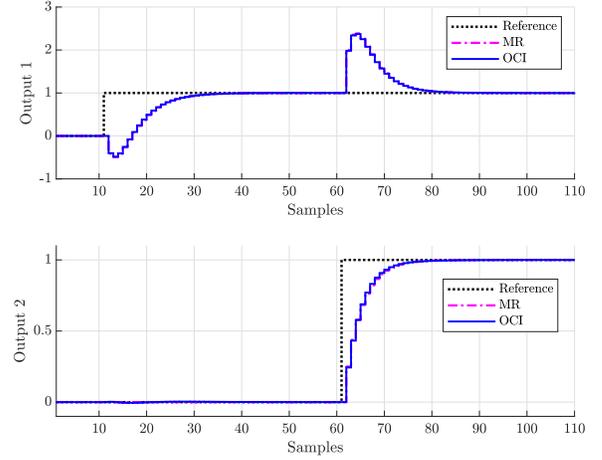}%
\caption{Closed-loop response with controller \eqref{eq:C2}.}
\label{fig:idealruido}%
\end{figure}

The resulting noise-free closed-loop response to step references is shown in Fig.~\ref{fig:idealruido}, where we have also plotted the reference model obtained\footnote{The effect of the noise on the outputs is omitted since the objective is to evaluate the reference tracking performance (see~\eqref{eq:JMR}).}.

Notice that \eqref{eq:Tdi2} is very similar to \eqref{eq:Td2} despite the low SNR of data and that the zero identified in $T_{d_{11}}(q,\hat \eta_N)$, $\hat z_{nm}=1.211$, is close to the actual value $z_{nm}=1.2$. Moreover, the response obtained with \eqref{eq:C2} is almost equal to the one provided by~\eqref{eq:Tdi2}. The cost~\eqref{eq:JMR} is $J^{MR}(\hat P_N,\hat\eta_N)=4\times 10^{-5}$. Notice also that since the coupling element of $T_{d}(q,\hat \eta_N)$ is identified in order to fulfill the zero output direction restriction, we do not have control on the effect of such coupling. In the example, the disturbance is of high amplitude. This is the price to pay in order to  eliminate entirely the nasty effect of the NMP from the (chosen as the most important) output $\#2$.

\begin{figure}[htb]%
\includegraphics[width=0.5\textwidth]{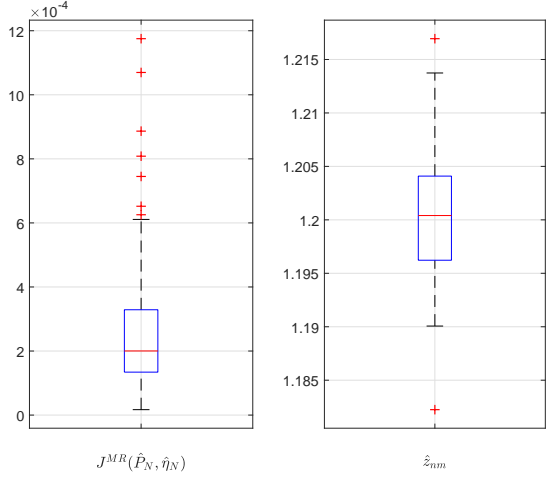}%
\caption{Box-plots of $J^{MR}(\hat P_N,\hat\eta_N)$ and $\hat z_{nm}$ for 100 Monte Carlo runs using the reference model given by \eqref{eq:td2eta}.}
\label{fig:MC2}%
\end{figure}

For completeness, we performed one hundred Monte Carlo runs of the same simulation modifying the realizations of the output noise and also of the PRBS input. Fig. \ref{fig:MC2} depicts the resulting box-plots\footnote{On each box, the central mark is the median, the edges of the box are the $25$th and $75$th percentiles, the whiskers extend to the most extreme data points not considered outliers, and outliers are plotted individually.} of $J^{MR}(\hat P_N,\hat\eta_N)$ and $\hat z_{nm}$, which show that the OCI method is specially suitable for cases where the SNR is relatively low, as already noted in \cite{Huff:Campestrini:Goncalves:Bazanella:2019}. In particular, the median of $\hat z_{nm}$ is practically equal to $z_{nm}$.

For comparison purposes, consider now the more realistic situation in which Assumption~\ref{asp:fullorder} is not satisfied. We choose a reference model faster than \eqref{eq:td2eta}:
\begin{equation} \label{eq:td3eta}
\hspace*{-0.2cm}\Tde = \begin{bmatrix}
 \frac{\eta_{1,11}q+0.16-\eta_{1,11}}{(q-0.6)^2} & \frac{(q+\eta_{2,12})(q-1)}{(q-0.6)^3} \\[4pt]
 0 & \frac{0.4}{(q-0.6)} \\[4pt]
\end{bmatrix}.
\end{equation}
Consider also colored noise at the system's outputs with $H_0(q)=\frac{q}{q-0.3}I$ in \eqref{eq:y(t)}, where $\Lambda$ is the same as before.

\begin{figure}[H]%
\includegraphics[width=0.5\textwidth]{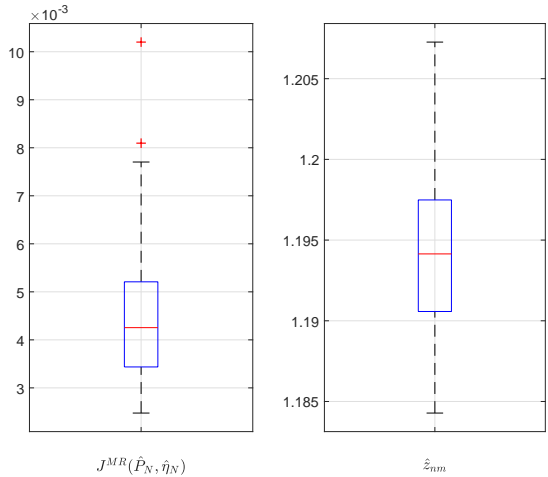}%
\caption{Box-plots of $J^{MR}(\hat P_N,\hat\eta_N)$ and $\hat z_{nm}$ for 100 Monte Carlo runs using the reference model given by \eqref{eq:td3eta}.}
\label{fig:MC3}%
\end{figure}

\begin{figure}[H]%
\includegraphics[width=0.5\textwidth]{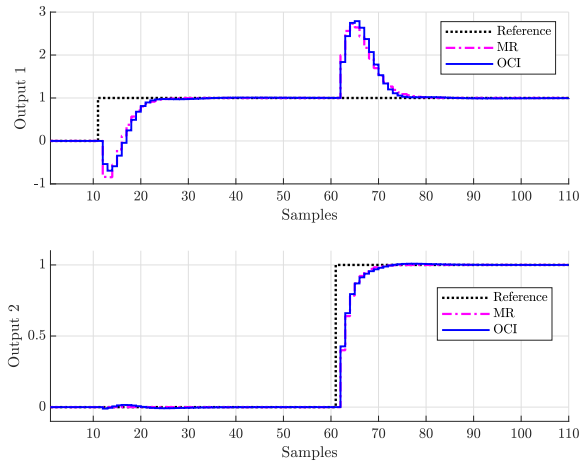}%
\caption{Closed-loop response with controller \eqref{eq:C3}.}
\label{fig:naoidealruidocolorido}%
\end{figure}

Fig.~\ref{fig:MC3} shows the resulting box-plots of $J^{MR}(\hat P_N,\hat\eta_N)$ and $\hat z_{nm}$, while Fig.~\ref{fig:naoidealruidocolorido} shows the resulting closed-loop response for a single realization of the simulation, where $C(q,\hat P_N)$ and $T_d(q,\hat\eta_N)$ are given, respectively, by
\begin{equation}
\hspace*{-0.2cm}\scalebox{0.8}{$C(q,\hat P_N)=$}
	\begin{bmatrix}
		\frac{0.83(q^2-1.734q+0.752)}{q(q-1)}&\frac{-0.45(q-0.918)(q-0.699)}{q(q-1)}\\
		\frac{-0.69(q^2-1.737q+0.755)}{q(q-1)}&\frac{0.65(q-0.823)(q-0.457)}{q(q-1)}
	\end{bmatrix}
	\label{eq:C3}
\end{equation}
\begin{equation}
T_d(q,\hat \eta_N)=
	\begin{bmatrix}
		\frac{-0.835(q-1.192)}{(q-0.6)^2}&\frac{(q-1)(q-0.243)}{(q-0.6)^3}\\
		0&\frac{0.4}{q-0.6}
	\end{bmatrix}
	\label{eq:Tdi3}
\end{equation}
and $J^{MR}(\hat P_N,\hat\eta_N) =  4\times 10^{-3}$.

Note that now the median of $\hat z_{nm}$ differs from $z_{nm}=1.2$ by a certain amount and that the MR cost $J^{MR}(\hat P_N,\hat\eta_N)$ is much higher than before (comparing the median values), which is to be expected from the fact that Assumption 1 is not satisfied. 

\section{Conclusions}
\label{sec:conclusions}

A flexible formulation of the OCI method was developed, which is specially suitable to deal with non-minimum phase systems, where the NMP transmission zeros of the plant are identified together with the controller parameters. In this formulation, the poles of the closed-loop system are allocated at prescribed positions, while the numerators of the reference model are parametrized and let free to vary. The presented methodology handles not only diagonal but more general reference model structures. 
In this work, we provided a case study in which the NMP effect is moved to a specific output through the use of a block-triangular reference model, identified together with the controller parameters.

Notice that the reference model structure can be customized by the user of the method. The diagonal and block-triangular ones presented in Section~\ref{sec:parametrization} are just illustrative examples. Moreover, it is worth saying that the flexible formulation of the method can, in principle, also be advantageous for systems that do \textit{not} have NMP transmission zeros, even if this case was not the focus of this work (see, for instance, \cite{Bordignon:Campestrini:2019} in the context of disturbance rejection).

As a final remark, note that the OCI method performs well even when data are affected by noise with low SNR, as shown by the simulation results and as discussed in \cite{Campestrini:Eckhard:Bazanella:Gevers:2017,Huff:Campestrini:Goncalves:Bazanella:2019,Huff:Goncalves:Campestrini:2018}.


\bibliographystyle{ieeetr}
\bibliography{referencias}

\end{document}